# Exchange bias and training effects in antiferromagnetically coupled $La_{0.7}Sr_{0.3}MnO_3$ / $SrRuO_3$ superlattices

S. Narayana Jammalamadaka[(a)], J Vanacken and V. V. Moshchalkov

*INPAC – Institute for Nanoscale Physics and Chemistry, K.U. Leuven, Celestijnenlaan 200D, B–3001 Leuven, Belgium*

**Abstract** :- Exchange bias (EB) and the training effects (TE) in an antiferromagnetically coupled $La_{0.7}Sr_{0.3}MnO_3$ / $SrRuO_3$ superlattices were studied in the temperature range 1.8 – 150 K. Strong antiferromagnetic (AFM) interlayer coupling is evidenced from AC - susceptibility measurements. Below 100 K, vertical magnetization shifts are present due to the two remanent states corresponding to the two ferromagnetic (FM) layers at FM and AFM coupling condition. After field cooling (FC), significant decrease in the exchange bias field ($H_{EB}$) is observed when cycling the system through several consecutive hysteresis loops. Quantitative analysis for the variation of $H_{EB}$ vs. number of field cycles ($n$) indicates an excellent agreement between the theory, based on triggered relaxation phenomena, and our experimental observations. Nevertheless, the crucial fitting parameter $K$ indicates smooth training effect upon repeated field cycling, in accordance with our observation.

PACS 75.70.Cn– Magnetic properties of interfaces

PACS 75.60.Ej – Magnetization curves, hysteresis

PACS 75.75.-c– Magnetic properties of nanostructures

The interfacial interactions in magnetic multilayers are crucial, as they can dramatically change the magnetic response of the overall structure. These interactions can be controlled by external parameters such as magnetic field and temperature. Shift in the magnetic hysteresis loop along the magnetic field axis would be evident upon tuning the external parameters and this 'shift' has been termed as "Exchange bias (EB) effect" [1 – 3]. EB effects have extensively been studied in magnetic systems consisting of ferromagnetic (FM) - antiferromagnetic (AFM) [1], FM - spin glass [4], AFM – ferrimagnetic [5] and FM – FM bilayers [6]. Among these, FM/FM bilayer systems are important in understanding core issues related to exchange bias such as the possible origins of the hysteresis loop asymmetry and training effects [7]. From the minor loop measurements, EB effects have also been observed in bilayers composed of FM layers with different magnitudes of magnetic anisotropy [7, 8]. The reduction in $H_{EB}$ upon repeated field cycling is termed as training effect of EB system and is manifested as $H_{EB}$ vs $n$, where $n$ is the number of field cycles after FC process [8, 9]. EB and training effects are important for potential applications in ultrahigh-density magnetic recording, giant magnetoresistance, and spin valve devices [10].

$La_{0.7}Sr_{0.3}MnO_3$ (LSM) / $SrRuO_3$ (SR) superlattices have gained much attention due to the antiferromagnetic interlayer coupling which depends sensitively on the magnetocrystalline anisotropy and interfacial intermixing [12]. $La_{0.7}Sr_{0.3}MnO_3$ (LSM) [13] and $SrRuO_3$ (SR) [14] have bulk ferromagnetic Curie temperatures of 370 K and 160 K, respectively. One can grow these superlattices epitaxially and can change magnetic properties by cation substitution [15]. In these superlattices, the soft layer $La_{0.7}Sr_{0.3}MnO_3$ is antiferromagnetically coupled with the hard $SrRuO_3$ pinning layer below 150 K. As the hard layer, $SrRuO_3$ is a FM, it gives unique opportunity to change its magnetization state with the magnetic field and in turn will allow studying the exchange bias and training effects based on the pinning layer magnetization configuration. Nevertheless, exchange bias effects in this interlayer coupled $La_{0.7}Sr_{0.3}MnO_3$/

[(a)]E-mail: Surya.Jammalamadaka@fys.kuleuven.be

SrRuO$_3$ superlattices have yet to be established. To understand the temperature dependence of EB and its thermal cycling effects, further sensitive experimental verification is required, in particular to validate possible spin - based applications. Hence, in this paper we study the exchange bias and training effects in the temperature range 1.8 – 150 K of La$_{0.7}$Sr$_{0.3}$MnO$_3$/SrRuO$_3$ superlattices. The intriguing observations that we would like to emphasize are vertical and horizontal loop shifts in the entire temperature range of investigation. This paper also describes the quantification of the training effect with the existing theoretical models.

The superlattice of La$_{0.7}$Sr$_{0.3}$MnO$_3$/SrRuO$_3$ was fabricated by pulsed laser deposition employing a KrF excimer laser. Substrate temperature and oxygen partial pressure were 650°C and 0.14 mbar respectively. SrTiO$_3$ (001) substrate with a miscut angle of about 0.1° was used. A total of 30 layers of La$_{0.7}$Sr$_{0.3}$MnO$_3$ and SrRuO$_3$ with thicknesses of 2.3 nm and 3.3 nm, respectively, were grown. The microstructure of the superlattice was investigated by X – ray diffraction, atomic force microscopy, high resolution transmission electron microscope and high angle annular dark field scanning transmission electron microscopy (HAADF-STEM) [15]. Figure 1 shows a HAADF-STEM micrograph of the La$_{0.7}$Sr$_{0.3}$MnO$_3$ / SrRuO$_3$ superlattice. The interfaces between the La$_{0.7}$Sr$_{0.3}$MnO$_3$ and SrRuO$_3$ layers are free of misfit dislocations. However, the interfacial atomic layers were affected by intermixing; both A-site (La/Sr) and B-site (Mn/Ru) cations intermix in 1-2 unit cells across the interface, as marked by the rectangles in Fig 1. Magnetic properties were measured using a vibrating sample magnetometer (Oxford instruments). AC-susceptibility studies were carried out using a Physical Property Measurement System (PPMS, Quantum design) in the temperature range of 1.8 - 300 K.

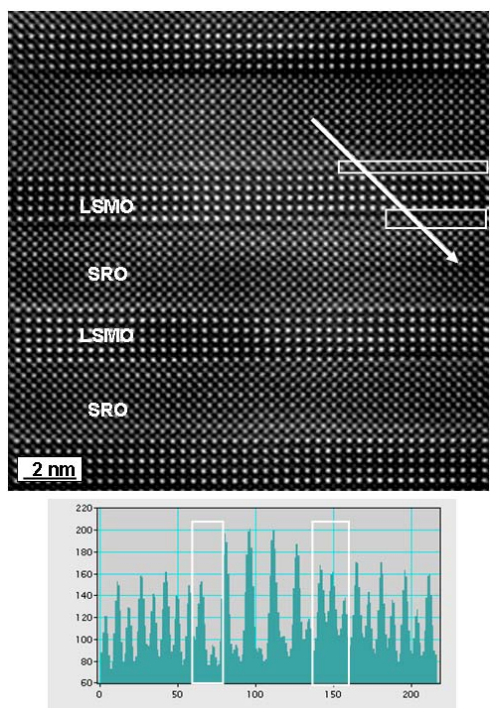

**Fig. 1**: (Color online) HAADF-STEM investigations of the La$_{0.7}$Sr$_{0.3}$MnO$_3$ / SrRuO$_3$ superlattices: Upper image shows a Z-contrast STEM micrograph and the lower graph plots the peak intensities along the white arrow, as indicated in the upper panel. The interfacial regions affected by intermixing of the A-site and B-site cations are marked by the white rectangles.

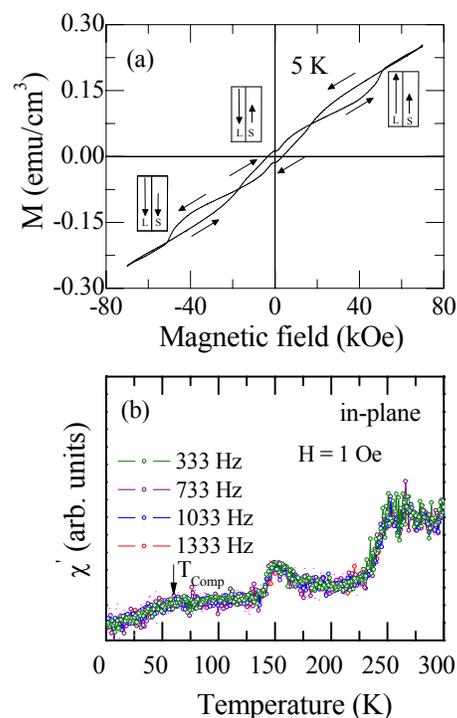

**Fig. 2**: (Color online) (a) Inverted hysteresis loop of La$_{0.7}$Sr$_{0.3}$MnO$_3$ / SrRuO$_3$ superlattice at 5 K (b) AC-Susceptibility graph at different frequencies in the temperature range 2 – 300 K.

In order to elucidate the existence of the crossing of hysteresis loops, initially we show isothermal magnetization behavior at 5 K in Fig. 2a. The data were collected while sweeping the magnetic field at a rate of 3 kOe/min. Starting at high fields, magnetization decreases gradually with magnetic field, and below 15 kOe, crossing of the central part of the loop is observed,[16] which can be explained as a result of AFM coupling and giant exchange bias. Fig. 2b shows the AC - susceptibility measurements on the $La_{0.7}Sr_{0.3}MnO_3$ / $SrRuO_3$ superlattices at various frequencies 333, 733, 1033 and 1333 Hz, respectively, in an alternating field of 1 Oe. Measurements were carried out with the field applied parallel to the superlattice surface (in-plane) and were performed after ZFC of the specimen from 300 K. Around 300 K, the value of susceptibility is almost constant as though there is a ferromagnetic transition at high temperature due to the $La_{0.7}Sr_{0.3}MnO_3$ layers [17]. Below 150 K, a decrease in the susceptibility is apparent, indicating a strong AFM interlayer coupling. The fact that the magnetic ordering does not arise from spin-glass freezing is evidenced by the observation that the peak in $\chi'$ is found to be frequency independent. At 62 K, a compensation point ($T_{comp}$) is clearly seen from AC – susceptibility, however, this compensation point is absent in the DC magnetization graphs [16].

A detailed study of the minor loops was performed to explore the exchange bias phenomenon and vertical loop shifts using the following protocol: Hysteresis loops were measured from -15 to 15 kOe at different temperatures 5, 25, 50, 75, 100, 125 and 150 K in ZFC and FC conditions (Fig. 3). For each FC measurement, the sample was warmed up to 300 K and cooled to the desired temperature in the presence of 10 kOe. The low field response of these minor loops is dominated by magnetization reversal of the $La_{0.7}Sr_{0.3}MnO_3$. The hysteresis-loop midpoints of the ZFC and FC curves differ;

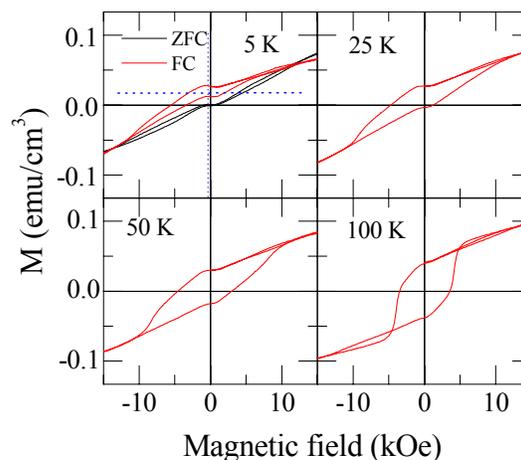

**Fig. 3**: (Color online) Zero field cooled (ZFC) and field cooled (FC) curves are shown at different temperatures. The horizontal as well as vertical shifts are clearly visible. The dotted line shows the new center of gravity of FC curve. For clarity the dotted lines are not shown for other temperatures.

as the vertical midpoint shift along the field axis is indicated by the dotted lines for the 5 K data (for clarity dotted lines are not shown for the other temperatures). The origin of this shift could be that below 150 K, ferromagnetic hard layer $SrRuO_3$ pins the magnetically soft layer, $La_{0.7}Sr_{0.3}MnO_3$ and shifts the hysteresis loops along the magnetic field axis. This shift can be quantified as $H_{EB}$.

Fig. 4 shows the $H_{EB}$ and coercivity ($H_C$) as a function of temperature for the $La_{0.7}Sr_{0.3}MnO_3$ / $SrRuO_3$ superlattices. $H_{EB}$ was determined from the horizontal shift in the midpoint of the minor loop. $H_{EB}$ decreases approximately linearly with increasing temperature in the low temperature region and goes to zero around the blocking temperature $T_B \sim 100$ K. It is also evident from the Fig. 3 that there is a vertical shift for the hysteresis loop, at all temperatures below 100 K. The presence of vertical loop shift below 100 K can be linked to the presence of two remanent states corresponding to the two FM layers at parallel and anti-parallel coupling condition. In contrast, above 100 K, the lack of magnetic anisotropy for SrRuO$_3$ layer indicates the absence of vertical loop shift. This also shows that the magnetization of the hard layer is fixed while that of the magnetization of the soft layer is switched. At this point it is worth mentioning that exchange bias and vertical loop shifts have been observed in CoCr/CoPtCrB ferromagnetic bilayers and these shifts have been explained on the basis of remanent magnetic moment of CoPtCrB layer [8]. In contrast, coercivity initially increased and then decreased with temperature as shown in Fig. 4.

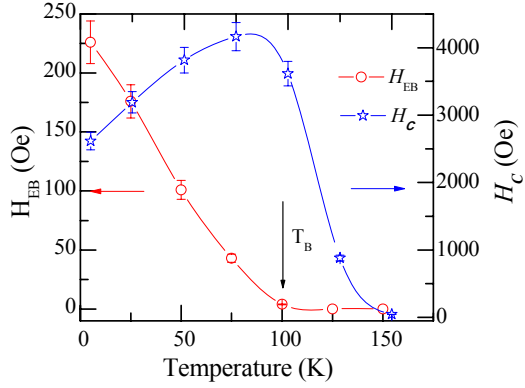

**Fig. 4**: (Color online) (a) Exchange bias field ($H_{EB}$) and coercivity ($H_C$) as a function of temperature. $H_{EB}$ decreases approximately linearly with increasing temperature in the low temperature region and gradually disappears around the blocking temperature $T_B \sim 100$ K.

Fig. 5 (a, b) illustrate the $H_{EB}$ and normalized remanent magnetization as a function of cooling field. With the cooling field, below 10 kOe, the increase in remnant magnetization ($M_r$) is apparent, suggesting that the sample's magnetization is higher for higher cooling field because of the presence of two remanent magnetic states. Concurrently, the increase in $H_{EB}$ also evident with the cooling field, as the effective Zeeman energy increases, and this can make more and more spins align until all spins are parallel to the external field, which results in increase in the $H_{EB}$. At higher fields ($\geq 10$ kOe), the pinning mechanism that exists between La$_{0.7}$Sr$_{0.3}$MnO$_3$ and SrRuO$_3$ vanishes and hence the decrease in $H_{EB}$ is evident as shown in Fig. 5a. Below 10 kOe, the Zeeman coupling is not strong enough to compete with the antiferromagnetic coupling which exists between La$_{0.7}$Sr$_{0.3}$MnO$_3$ and SrRuO$_3$. Hence 10 kOe can be considered as an effective de-pinning threshold field above which the magnetic interactions are overcome by the Zeeman coupling. Above this field, due to strong Zeeman coupling, the La$_{0.7}$Sr$_{0.3}$MnO$_3$ and SrRuO$_3$ layers less strongly coupled and thus exerts a weaker

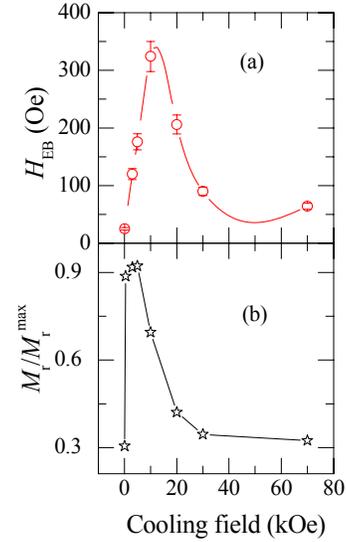

**Fig. 5:** (Color online) (a) Exchange bias field ($H_{EB}$) and (b) normalized remanent magnetization ($M_r/M_r^{max}$) as a function of cooling field ($H_{cool}$).

pinning, resulting in a decrease in the $H_{EB}$ and remnant magnetization.

Training effects are interesting characteristics of EB phenomena and are caused by the non-equilibrium nature of the spin structure in the pinning layer. Training effect, which is manifested as the gradual decrease in $H_{EB}$ when cycling the system through several consecutive hysteresis loops, is a clear indication of rearrangements in the pinning layer spin structure towards an equlibrium configuration [8, 18, 19]. In order to investigate the training effect of the exchange bias, a series of hysteresis loops were consecutively measured at 5 K after FC. The ZFC, first, second and third FC loops are shown in the mainframe of Fig. 6. The inset of Fig. 6 shows the $H_{EB}$ vs. $n$. Upon closer observation, the training effect is very strong for low $n$ values and becomes less pronounced for higher $n$, which indicates that the related relaxation processes in biasing hard layer occur predominantly during the first few reversals, while subsequent loops produce only minor changes; nevertheless, training effect in the present system indicates a smooth variation.

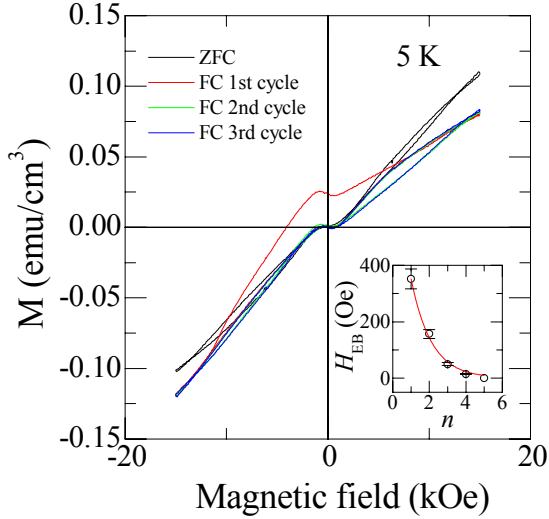

**Fig. 6:** (Color online) Minor hysteresis loops of ZFC and FC. FC curves were measured five times consecutively after ZFC to see the thermal cycling effect. Inset shows the quantification of the training effect to the theory based on triggered relaxation phenomena. The inset shows the variation of $H_{EB}$ with the number of field cycles after FC. The scattered symbols are experimental data and solid line shows the theoretical fit.

Discretized dynamical Landau – Khalatnikov equation is used in order to address the training effects in $La_{0.7}Sr_{0.3}MnO_3$ / $SrRuO_3$ superlattices [19, 20]. Basically this approach has been proposed to calculate the training effect in FM bilayer system [8]. However, we could well explain our experimental observations using equation (1). The explicit expression for the cycle dependent bias field is

$$H_{EB}(n) = (K+1)^{n-1}\left\{H_{EB}(1) - KH_{EB}^e\left[\frac{(K+1)^{n+1}-1}{K(K+1)^{n-1}} - (K+2)\right]\right\} \quad (1)$$

where $K$ is the crucial fitting parameter quantifying the rate of change in $H_{EB}(n)$, where $n$ is number of field cycles. The parameter, $H_{EB}(1)$ is the first point in the $H_{EB}$ vs $n$ data and $H_{EB}^e$ is the equilibrium bias field which indicates the $H_{EB}$ value when $n$ approaches ∞. $K$ and $H_{EB}^e$ enter the above equation as fitting parameters and $H_{EB}(n=1)$ is a fixed value. The bottom inset of Fig. 6 shows the variation of $H_{EB}$ vs $n$. The scattered points correspond to the experimental data and the solid red line shows the result of least square fitting of the equation (1). There is an excellent agreement between the theory and experiment; $K$ has a value of -0.59. The value of $K$ is within the range $-1 \leq K \leq 0$, as it has been observed for other systems [18]. The explicit function in equation (1) can be written as

$$H_{EB}(n+1) = (K+1)H_{EB}(n) - KH_{EB}^e \quad (2)$$

If the value of $K = 0$, $H_{EB}(n+1) = H_{EB}(n)$ which means that no training effect in the system. Similary, $K = -1$ implies that a step like change in the bias field between first two points and zero training for $n > 2$. In our case extracted value of $K$ is -0.59, which indicates that there is a TE and a smooth relaxation process in biasing the hard layer for the number of field cycles that were investigated. In addition, the value of $K$ also indicates that the hard layer which is not in equilibrium reaches equilibrium state smoothly upon repeated field cycling.

In summary, AC - susceptibility and DC – magnetization were used to probe the origin of exchange bias and training effects in an antiferromagnetically coupled $La_{0.7}Sr_{0.3}MnO_3$ / $SrRuO_3$ superlattices. The remanent states corresponding to the two FM layers at parallel and anti-parallel coupling condition below 100 K are responsible for the observed vertical loop shifts. The reduction in $H_{EB}$ is observed upon repeated field cycling. These training effects have been explained by using theoretical approach based on the discretized dynamical Landau − Khalatnikov equation. Quantitatively, there is an excellent agreement between the theory and experimental data, indicating that the origin of the training effect is related to the relaxation mechanism of pinning layer spin structure towards an equilibrium configuration. The value of $K$ is -0.59, which indicates smooth training effect upon repeated field cycling.

SNJ would like to thank KU Leuven, for research fellowship. This work is supported by the K.U. Leuven Excellence financing (INPAC), by the Flemish Methusalem financing and by the IAP network of the Belgian Government. SNJ would also like to thank Dr. Ionela Vrejoiu and Dr. Eckhard Pippel from Max-Planck-Institut für Mikrostrukturphysik (MPI) − Halle for the sample and STEM investigation of the sample.